\documentclass{svmult6}
\usepackage{graphicx}
\usepackage[abbr]{harvard}
\newcommand{\mk}{{\mathbf k}}
\newcommand{\mx}{{\mathbf x}}

\newcommand{\beq}{\begin{equation}}
\newcommand{\eeq}{\end{equation}}
\newcommand{\realR}{{{{\rm I\kern-0.16em{}R}}}}
\begin{document}
\pagenumbering{arabic}
\chapter{Statistics of Galaxy Clustering}
\chapterauthors{Vicent J. Mart\'{\i}nez\footnote{Observatori Astron\`omic, 
Universitat de Val\`encia, Burjassot, 46100
Val\`encia, Spain, e-mail: vicent.martinez@uv.es}\\
Enn Saar\footnote{Tartu Observatoorium, T\~oravere, 61602, Estonia,
e-mail: saar@aai.ee}}
\shortauthname{Vicent J. Mart\'{\i}nez and Enn Saar}
\begin{abstract}
In this introductory talk we will establish connections
between the statistical analysis of galaxy clustering
in cosmology and recent work in mainstream spatial statistics.
The lecture will review the methods of spatial statistics used by 
both sets of scholars, having in mind the cross-fertilizing purpose 
of the meeting series.
Special topics will be: description of the galaxy samples, selection
effects and biases, correlation functions, 
nearest neighbor distances, void probability functions,
Fourier analysis, and structure statistics. 
\end{abstract} 
\section{Introduction}
One of the most important motivations of these series of conferences
is to promote vigorous interaction between statisticians and astronomers.
The organizers merit our admiration for bringing together such a stellar 
cast of colleagues from both fields. In this third edition, one of the
central subjects is cosmology, and in particular, statistical 
analysis of the large-scale structure in the universe. There is a 
reason for that --- the rapid increase of the amount and quality of the 
available observational data on the galaxy distribution (also on 
clusters of galaxies and quasars) and
on the temperature fluctuations of the microwave background radiation.

These are the two fossils of the early universe on which cosmology,
a science driven by observations, relies.
Here we will focus on one of them --- the galaxy distribution.
First we briefly review the redshift surveys, how they are built and
how to extract statistically analyzable samples from them, considering
selection effects and biases. Most of the statistical analysis
of the galaxy distribution are based on second order methods (correlation 
functions and power spectra). We comment them, providing the connection
between statistics and estimators used in cosmology and in spatial
statistics. Special attention is devoted to the analysis of clustering in
Fourier space, with new techniques for estimating the power spectrum,
which are becoming increasingly popular in cosmology. 
We show also the results of applying these second-order methods 
to recent galaxy redshift surveys.

Fractal analysis has become very popular as a consequence of the scale-invariance
of the galaxy distribution at small scales, reflected in the power-law 
shape of the two-point correlation function. We discuss here some 
of these methods and the results of their application to the observations,
supporting a gradual transition from a small-scale fractal regime to 
large-scale homogeneity. The concept of lacunarity is illustrated
with some detail. 

We end by briefly reviewing some of the alternative measures of point 
statistics and structure functions applied thus far to the galaxy
distribution: void probability
functions, counts-in-cells, nearest neighbor distances,
genus, and Minkowski functionals.

\section{Cosmological datasets}

Cosmological datasets differ in several respects from those usually
studied in spatial statistics. The point sets in
cosmology (galaxy and cluster surveys) bear the imprint of
the observational methods used to obtain them.

The main difference is the 
systematically variable intensity (mean density) of
cosmological surveys. These surveys are usually magnitude-limited,
meaning that all objects,
which are brighter than a pre-determined limit,
are observed in a selected region of the sky.
This limit is mainly determined by the telescope and 
other instruments used for the program. Apparent
magnitude, used to describe the limit,
is a logarithmic measure of the observed radiation
flux.

It is usually assumed that galaxies at all distances have the same 
(universal) luminosity distribution function.
This assumption has been tested and found to be in
satisfying accordance with observations. As the observed
flux from a galaxy is inversely proportional to the square of its distance,
we can see at larger distances only a bright fraction
of all galaxies. This leads directly to the mean density
of galaxies that depends on their distance from us $r$.

This behaviour is quantified by a selection function $\phi(r)$,
which is usually found by estimating first
the luminosity distribution of galaxies (the luminosity
function).

One can also select a distance limit, find the minimum
luminosity of a galaxy, which can yet be seen at that distance,
and ignore all galaxies that are less luminous.
Such samples are called volume-limited. They are used
for some special studies (typically for counts-in-cells), but
the loss of hard-earned information is enormous. 
The number of galaxies in volume-limited samples is 
several times smaller than in the parent magnitude-limited
samples. This will also increase the shot (discreteness) 
noise.

In addition to the radial selection function $\phi(r)$,
galaxy samples also are frequently subject to angular
selection. This is due to our position in the Galaxy ---
we are located in a dusty plane of the Galaxy, and
the window in which we see the Universe, also is
dusty. This dust absorbs part of galaxies' light,
and makes the real brightness limit of a survey
dependent on the amount of dust in a particular 
line-of-sight. This effect has been described by a
$\phi(b)\sim(\sin b)^{-1}$ law ($b$ is the galactic
latitude); in reality the dust
absorption in the Galaxy is rather inhomogeneous.
There are good maps of the amount of Galactic dust in the sky, 
the latest maps have been obtained
using the COBE and IRAS satellite data 
\cite{dust}.

Edge problems, which usually affect estimators in
spatial statistics, also are different for cosmological
samples. The decrease of the mean density towards the sample
borders alleviates these problems. Of course, if we
select a volume-limited sample, we select also all these
troubles (and larger shot noise). From the 
other side, edge effects are made more prominent by
the usual observing strategies, when surveys are
conducted in well-defined regions in the sky.
Thus, edge problems are only partly alleviated; 
maybe it will pay to taper our samples at the side
borders, too?

Some of the cosmological surveys have naturally
soft borders. These are the all-sky surveys; the
best known is the IRAS infrared survey, dust is
almost transparent in infrared light. The corresponding
redshift survey is the PSCz survey, which covers about
85\% of the sky \cite{pscz}. A special follow-up survey is in progress
to fill in the remaining Galactic Zone-of-Avoidance region,
and meanwhile numerical methods have been developed to
interpolate the structures seen in the survey into the gap
\cite{vs,ballinger}.

Another peculiarity of galaxy surveys is that
we can measure exactly only the direction to the
galaxy (its position in the sky), but not its distance.
We measure the radial velocity $v_r$ (or redshift $z=v_r/c$,
$c$ is the velocity of light) of a galaxy, which
is a sum of the Hubble expansion, proportional to the
distance $d$, and the dynamical velocity $v_p$ of the galaxy,
$v_r=H_0d+v_p$. 
Thus we are differentiating between redshift space,
if the distances simply are determined as $d=v_r/H_0$,
and real space. The real space positions of galaxies could
be calculated if we exactly knew the peculiar velocities
of galaxies; we do not. The velocity distortions can be
severe; well-known features of redshift space are
fingers-of-God, elongated structures that are caused by
a large radial velocity dispersion in massive clusters of
galaxies. The velocity distortions expand a cluster in
redshift space in the radial direction five-ten times.

For large-scale structures the situation is different,
redshift distortions compress them. This is due to
the continuing gravitational growth of structures.
These differences can best be seen by comparing the
results of numerical simulations, where we know also
the real-space situation, in redshift space and in
real space.

The last specific feature of the cosmology datasets is
their size. Up to recent years most of the datasets have
been rather small, of the order of $10^3$ objects; exceptions
exist, but these are recent. Such a small number of 
points gives a very sparse coverage of three-dimensional
survey volumes, and shot noise has been a severe problem.

This situation is about to change, swinging to the other
extreme; the membership of new redshift surveys already
is measured in terms of $10^5$ (160,000 for the 2dF survey,
quarter of a million planned) and million-galaxy surveys
are on their way (the Sloan Survey). More information
about these surveys can be found in their Web pages:
\emph{http:/\kern-2pt/www.mso.anu.edu.au/\break
2dFGRS/\,} for
the 2dF survey and
\emph{http:/\kern-2pt/www.sdss.org/\,} for the Sloan survey.
This huge amount of data will force us to change 
the statistical methods we use. Nevertheless, the deepest
surveys (e.g., distant galaxy cluster surveys)
will always be sparse, so discovering small signals
from shot-noise dominated data will remain a necessary art.

\section{Correlation analysis}
There are several related quantities that are second-order
characteristics used to quantify clustering of the galaxy
distribution in real or redshift space. The most popular one in
cosmology is the two-point correlation function, $\xi({\mathbf r})$. The
infinitesimal interpretation of this quantity reads as follows:
\begin{equation}
\label{xi} dP_{12} = \bar{n}^2[ 1 + \xi({\mathbf r})]dV_1 dV_2
\end{equation}
is the joint probability that in each one of the two infinitesimal
volumes $dV_1$ and $dV_2$, with separation vector
$\mathbf r$, lies a galaxy. Here $\bar{n}$ is the mean
number density (intensity). Assuming that the galaxy distribution is a
homogeneous (invariant under translations) and isotropic
(invariant under rotations) point process, this probability depends
only on $r=|{\mathbf r}|$. In
spatial statistics, other functions related with $\xi(r)$ are
commonly used:
\begin{equation}
\lambda_2(r)=\bar{n}^2\xi(r)+1,  \qquad g(r)= 1+\xi(r), \qquad
\Gamma(r)= \bar{n}(\xi(r)+1),
\label{relxi}
\end{equation}
where $\lambda_2(r)$ is the second-order intensity function, $g(r)$ is the
pair correlation function, also called the radial distribution
function or structure function, and $\Gamma(r)$ is the conditional
density proposed by \citeasnoun{pietro87}.

Different estimators of $\xi(r)$ have been proposed so far in the
literature, both in cosmology and in spatial statistics. The main
differences are in correction for 
edge effects. Comparison of their performance can be found in
several papers
\cite{pons,kerm,stoyan}.
There is clear evidence that $\xi(r)$ is well
described by a power-law at scales $0.1 \leq r \leq 10 \, h^{-1}$ Mpc
where $h$ is the Hubble constant in units of 100 km s$^{-1}\,$Mpc$^{-1}$:
\[
\xi(r)=\left ( \frac{r}{r_0} \right )^{-\gamma},
\]
with $\gamma \simeq 1.8$ and $r_0 \simeq 5.4 \, h^{-1}$ Mpc. This
scaling behavior is one of the reasons that have lead some
astronomers to describe the galaxy distribution as fractal.
A power-law fit for $g(r) \propto r^{3-D_2}$ permits to
define the correlation dimension $D_2$. The extent of the fractal
regime is still a matter of debate in cosmology, but it seems
clear that the available data on redshift surveys indicate a
gradual transition to homogeneity  for scales larger than 15--20
$h^{-1}$ Mpc \cite{mart99}.
Moreover, in a fractal point
distribution, the correlation length $r_0$ increases with the
radius of the sample because the mean density decreases
\cite{pietro87}.
This simple prediction of the fractal
interpretation is not supported by the data, instead $r_0$ remains
constant for volume-limited samples with increasing depth
\cite{mart01}.

Several versions of the volume integral of the correlation
function are also frequently used in the analysis of galaxy
clustering. The most extended one in spatial statistics is the
so-called Ripley $K$-function
\begin{equation}
\label{rip}
K(r)= \int_0^r 4 \pi s^2 (1 + \xi(s)) ds
\end{equation}
although in cosmology it is more frequent to use an expression
which provides directly the average number of neighbors an
arbitrarily chosen galaxy has within a distance $r$,
$N(<r)=\bar{n}K(r)$ or the average conditional density
\[
\Gamma^{\ast}(r) = \frac{3}{r^3}\int_0^r \Gamma(s)s^2ds
\]
Again a whole collection of estimators are used to properly
evaluate these quantities. Pietronero and coworkers recommend to
use only minus--estimators to avoid any assumption regarding the
homogeneity of the process. In these estimators, averages
of the number of neighbors within a given distance are taken only
considering as centers these galaxies whose distances to the border are
larger than $r$. However, caution has to be exercised with this
procedure, because at large scales only a small number of centers
remain, and thus the variance of the estimator increases.

Integral quantities are less noisy than the corresponding
differential expressions, but obviously they do contain less
information on the clustering process due the fact that values of
$K(r_1)$ and $K(r_2)$ for two different scales $r_1$ and $r_2$ are
more strongly correlated than values of $\xi(r_1)$ and $\xi(r_2)$.
Scaling of $N(<r) \propto r^{D_2}$ provides a
smoother estimation of the correlation dimension. If scaling
is detected for partition sums defined by the moments of order $q$
of the number of neighbors
\[
Z(q,r)=\frac{1}{N}\sum_{i=1}^N n_i(r)^{q-1} \propto r^{D_q/(q-1)},
\]
the exponents $D_q$ are the so-called generalized or multifractal
dimensions \cite{mart90}.
Note that for $q=2$,
$Z(2,r)$ is an estimator of $N(<r)$ and therefore $D_q$ for $q=2$
is simply the correlation dimension. If different kinds of cosmic
objects are identified as peaks of the continuous matter density
field at different thresholds, we can study the correlation
dimension associated to each kind of object. The multiscaling
approach \cite{jensen} associated to the multifractal
formalism provides a unified framework to analyze this variation. It
has been shown \cite{mart95}
that the value of $D_2$
corresponding to rich galaxy clusters (high peaks of the density
field) is smaller than the value corresponding to galaxies (within
the same scale range) as prescribed in the multiscaling approach.

\begin{figure}
\begin{center}
    \leavevmode
    \includegraphics*[width=\textwidth]{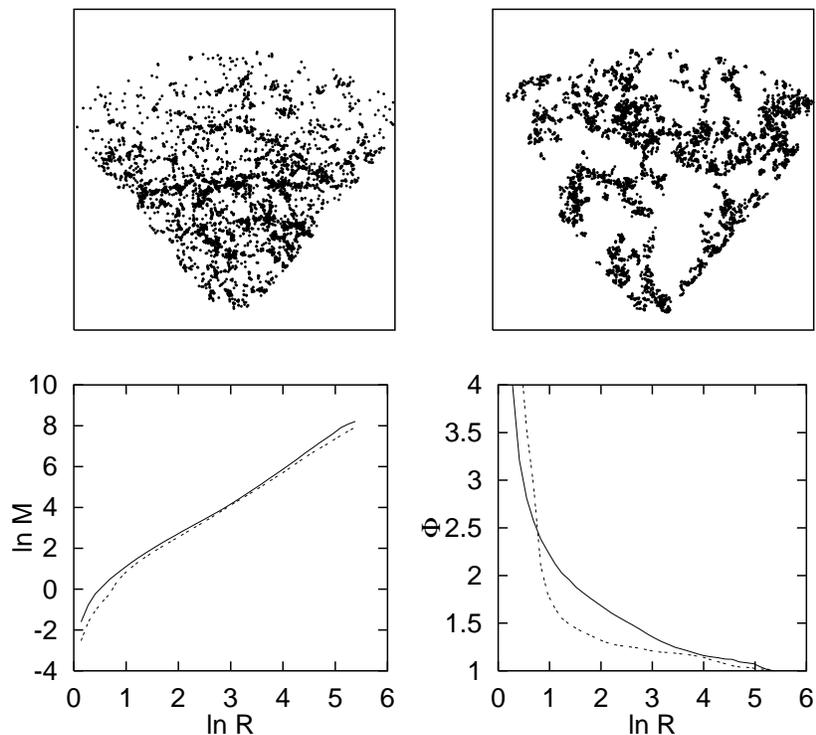}
  \end{center}
  \caption{Comparison of a Las Campanas survey slice (upper
  left panel) with the Rayleigh-L\'evy flight model (upper right panel).
  The fractal dimensions of both distributions coincide,
  as shown by the $\ln M$--$\ln R$ curves in the lower
  left panel, but the lacunarity curves (in the lower
  right panel) differ considerably. The solid lines describe
  the galaxy distribution, dotted lines -- the model
  results. 
  From {\protect\citeasnoun{msbook}}.
  }
\label{lacun}
\end{figure}

Finally we want to consider the role of lacunarity in the
description of the galaxy clustering \cite{msbook}.
In Fig. \ref{lacun}, we show the space distribution of 
galaxies within one slice of the Las Campanas redshift survey,
together with a fractal pattern generated by means of a
Rayleigh-L\'evy flight \cite{mandel}.
Both have the same
mass-radius dimension, defined as the exponent of the power-law
that fits the variation of mass within concentric spheres centered
at the observer position. 
\begin{equation}
M(R)=FR^{D_M}.
 \label{mrr}
\end{equation}
The best fitted value for both point
distributions is $D_M \simeq 1.6$ as shown in the left
bottom panel of Fig. \ref{lacun}.
The different appearance of both point distributions is a
consequence of the different degree of lacunarity. 
\citeasnoun{lacunar} have proposed to quantify this effect by
measuring the variability of the prefactor $F$ in Eq. \ref{mrr},
\[
\Phi = \frac{ E \{(F-\bar{F})^2 \}}{\bar{F}^2}
\]
The result of applying this lacunarity measure is shown in the
right bottom panel of Fig. \ref{lacun}. The visual differences between
the point distributions are now well reflected in this curve.

\section{Power spectra}

The current statistical model for the main cosmological fields
(density, velocity, gravitational potential) is the Gaussian 
random field. This field is determined either by
its correlation function or by its spectral density,
and one of the main goals of spatial statistics in
cosmology is to estimate those two functions.

In recent years the power spectrum
has attracted more attention than the
correlation function.
There are at
least two reasons for that --- the power spectrum
is more intuitive physically, separating processes
on different scales, and the model predictions
are made in terms of power spectra.
Statistically, the advantage is that the power spectrum
amplitudes for different wavenumbers are statistically
orthogonal: 
\[
E\left\{\widetilde{\delta}(\mk)\widetilde{\delta}^\star(\mk')\right\}=
    (2\pi)^3\delta_D(\mk-\mk')P(\mk).
\]
Here $\widetilde{\delta}(\mk)$ is the Fourier amplitude of
the overdensity field $\delta=(\rho-\bar{\rho})/\bar{\rho}$
at a wavenumber $\mk$, $\rho$ is the matter density,
a star denotes complex conjugation, $E\{\}$
denotes expectation values over realizations of the
random field, and $\delta_D(\mx)$ is the three-dimensional
Dirac delta function.
The power spectrum $P(\mk)$ is the Fourier
transform of the correlation function $\xi(\mathbf{r})$ of
the field. 

    Estimation of power spectra from observations
is a rather difficult task. Up to now the problem
has been in the scarcity of data; in the near future there
will be the opposite problem of managing huge data sets.
The development of statistical techniques here has been motivated 
largely by the analysis of CMB power spectra, where better
data were obtained first, and has been parallel to that recently.

The first methods developed to estimate the power
spectra were direct methods --- a suitable statistic was
chosen and determined from observations. A good reference
is \citeasnoun{fkp}.

The observed samples can be modeled by an inhomogeneous
point process (a Gaussian Cox process) of number density
$n(\mx)$: 
\[
n(\mx)=\sum_i\delta_D(\mx-\mx_i),
\]
where $\delta_D(\mx)$ is the Dirac delta-function.
As galaxy samples 
frequently have systematic density trends caused by
selection effects, 
we have to write the estimator of the density contrast
in a sample as
\[
D(\mx)=\sum_i\frac{\delta_D(\mx-\mx_i)}{\bar{n}(\mx_i)}-1,
\]
where $\bar{n}(\mx)\sim\bar{\rho}(\mx)$ is the selection
function expressed in the number density of objects.

The estimator for a Fourier amplitude 
(for a finite set of frequencies $\mk_i$) is
\[
F(\mk_i)=\sum_j\frac{\psi(\mx_j)}
    {\bar{n}({\mx}_j)}e^{i\mk_i\cdot\mx} -\widetilde{\psi}(\mk_i),
\]
where $\psi(\mx)$ is a weight function that can be selected
at will.
The raw estimator for the spectrum is
\[
P_R(\mk_i)=F(\mk_i)F^\star(\mk_i),
\]
and its expectation value
\[
E\left\{\langle|F(\mk_i)|^2\rangle\right\}
    =\int G(\mk_i-\mk')P(\mk')\,\frac{d^3k'}{(2\pi)^3}
    +\int_V\frac{\psi^2(\mx)}{\bar{n}(\mx)}\,d^3x,
\]
where $G(\mk)=|\tilde{\psi}(\mk)|^2$ is the window 
function that also depends on the geometry of the
sample volume.
Symbolically, we can get the estimate of the
power spectra $\widehat{P}$ by inverting the integral equation
\[
G\otimes \widehat{P}=P_R-N,
\]
where $\otimes$ denotes convolution, 
\index{power spectra!estimate}
$P_R$ is the raw
estimate of power, and $N$ is the (constant) shot noise term.

    In general, we have to deconvolve the noise-corrected
raw power to get the estimate of the power spectrum.
This introduces correlations in the estimated amplitudes, 
so these are not statistically orthogonal any more. 
A sample of a characteristic spatial
size $L$ creates a window function of width of $\Delta k\approx 1/L$,
correlating estimates of spectra at that wavenumber interval.

As the cosmological spectra are usually assumed to
be isotropic, the standard method to estimate the spectrum 
involves an additional step of averaging the estimates
$\widehat{P}(\mk)$ over a spherical shell 
$k\in[k_i,k_{i+1}]$ of thickness $k_{i+1}-k_i> \Delta k=1/L$
in  wavenumber space. 
The minimum-variance requirement gives the 
FKP \cite{fkp}
weight function:
\[
\psi(\mx)\sim\frac{\bar{n}(\mx)}{1+\bar{n}(\mx)P(k)},
\]
and the variance is
\[
\frac{\sigma^2_P(k)}{P^2_R(k)}\approx \frac{2}{\mathcal N},
\]
where $\mathcal N$ is the number of coherence
volumes in the shell. The number of independent volumes is
twice as small (the density field is real).
The coherence volume is 
$V_c(k)\approx(\Delta k)^3\approx 1/L^3\approx 1/V$.

As the data sets get large, straight application
of direct methods (especially the error analysis)
becomes difficult. There are different recipes that have
been developed with the future data sets in mind.
A good review of these methods is given in \citeasnoun{future}.

\begin{figure}
\begin{center}
    \leavevmode
    \includegraphics*[width=.8\textwidth]{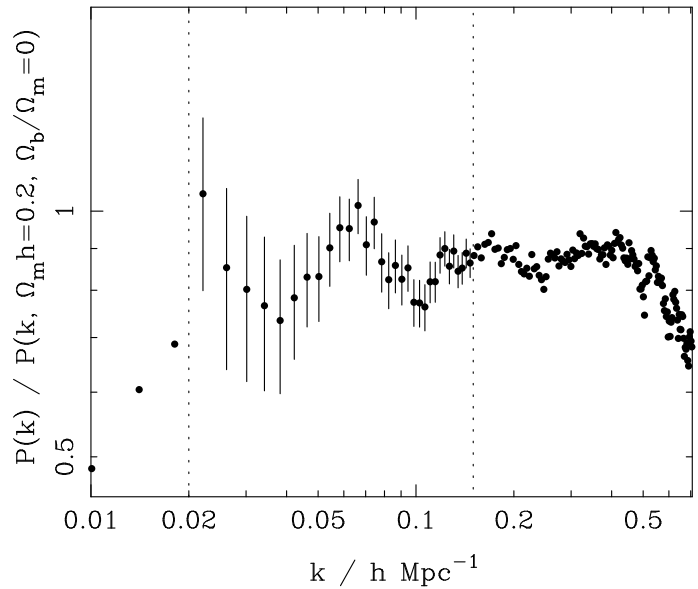}
  \end{center}
  \caption{         Power spectrum of the 2dF redshift survey,
         divided by a smooth model power spectrum.
         The spectrum is not deconvolved. Error bars are
         determined from Gaussian realizations; the dotted
         lines show the wavenumber region that is free
         of the influence of the window function and of
         the radial velocity distortions and nonlinear effects.
         (Courtesy of W. J. Percival and the 2dF galaxy redshift
         survey team.) }
\label{2dfpower}
\end{figure}

The deeper the galaxy sample, the smaller the coherence
volume, the larger the spectral resolution and the larger
the wavenumber interval where the power spectrum can be
estimated. The deepest redshift surveys presently available are the
PSCz galaxy redshift survey (15411 redshifts
up to about $400 h^{-1}\,$Mpc, see \citeasnoun{pscz}),
the Abell/ACO rich galaxy cluster survey, 637 redshifts
up to about 300$\,h^{-1}\,$Mpc \cite{miller}),
and the ongoing 2dF galaxy redshift survey (141400 redshifts
up to $750 h^{-1}\,$Mpc \cite{2dfnature}).
The estimates of power spectra for the two latter samples
have been obtained by the direct method 
\cite{batwig,2dfpower}.
Fig.~\ref{2dfpower} shows the power spectrum for
the 2dF survey.

The covariance matrix of the power spectrum estimates in Fig.~\ref{2dfpower}
was found from simulations of a matching Gaussian Cox
process in the sample volume. The main new feature in the spectra,
obtained for the new deep samples,
is the emergence of details (wiggles) in the power spectrum.
While sometime ago the main problem was to estimate the 
mean behaviour of the spectrum and to find its maximum,
now the data enables us to see and study the
details of the spectrum. These details have been
interpreted as traces of acoustic oscillations
in the post-recombination power spectrum. 
Similar oscillations are
predicted for the cosmic microwave background radiation
fluctuation spectrum. The CMB wiggles match the theory rather well, 
but the galaxy wiggles do not, yet.

Thus, the measurement of the power spectrum of the
galaxy distribution is passing from the determination of
its overall behaviour to the discovery and interpretation
of spectral details.

\section{Other clustering measures}
To end this review we briefly mention other measures used to
describe the galaxy distribution.

\subsection{Counts-in-cells and void probability function} 
The probability that a randomly placed sphere of radius $r$ contains
exactly $N$ galaxies is denoted by $P(N,r)$. In particular, for
$N=0$, $P(0,r)$ is the so-called void probability function,
related with the empty space function or contact distribution
function $F(r)$, more frequently used in the field of spatial
statistics, by $F(r)=1-P(0,r)$. The moments of the counts-in-cells
probabilities can be related both with the multifractal analysis
\cite{borgani93} and with the higher order $n$-point correlation
functions \cite{white79,stoy95,statstat}.

\subsection{Nearest-neighbor distributions} 
In spatial statistics,
different quantities based on distances to nearest neighbors have
been introduced to describe the statistical properties of point
processes. $G(r)$ is the distribution function of the distance $r$
of a given point to its nearest neighbor. It is interesting to
note that $F(r)$ is just the distribution function of the distance
$r$ from an arbitrarily chosen point in $\realR^3$ --- not being an event
of the point process --- to a point of the point process (a galaxy
in the sample in our case). The quotient
\[
J(r) = \frac{1-G(r)}{1-F(r)}
\]
introduced by
\citeasnoun{lieshout}
is a powerful tool
to analyze point patterns and has discriminative power to compare
the results of $N$-body models for structure formation with the
real distribution of galaxies \cite{kerscher99}.

\subsection{Topology} 
One very popular tool for analysis of the galaxy
distribution is the genus of the isodensity surfaces. To define
this quantity, the point process is smoothed to obtain a
continuous density field, the intensity function, by means of a
kernel estimator for a given bandwidth. Then we consider the
fraction of the volume $f$ which encompasses those regions having
density exceeding a given threshold $\rho_t$. The boundary of these regions
specifies an isodensity surface. The genus $G(S)$ of a surface $S$ is
basically the number of holes minus the number of isolated regions
plus 1. The genus curve shows the variation of $G(S)$ with $f$ or $\rho_t$ 
for a given window radius of the kernel function. An analytical
expression for this curve is known for Gaussian density fields. 
It seems that the
empirical curve calculated from the galaxy catalogs can be
reasonably well fitted to a Gaussian genus curve \cite{canavezes}
for window radii varying within a large range of scales.

\subsection{Minkowski functionals} 
A very elegant generalization of the
previous analysis to a larger family of morphological
characteristics of the point processes is provided by the
Minkowski functionals. These scalar quantities are useful to study
the shape and connectivity of  a union of convex bodies. They are
well known in spatial statistics and have been introduced in
cosmology by 
\citeasnoun{mecke94}.
On a clustered
point process, Minkowski functionals are calculated by
generalizing the Boolean grain model into the so-called germ-grain
model. This coverage process consists in considering the sets $A_r
= \cup_{i=1}^N B_r({\mathbf x}_i)$ for the diagnostic parameter
$r$, where $\{ {\mathbf x}_i \}_{i=1}^N$ represents the galaxy
positions and $B_r({\mathbf x}_i)$ is a ball of radius $r$
centered at point ${\mathbf x}_i$. Minkowski functionals are
applied to sets $A_r$ when $r$ varies. In $\realR^3$ there are
four functionals: the volume $V$, the surface area $A$, the
integral mean curvature $H$, and the Euler-Poincar\'e
characteristic $\chi$, related with the genus of the boundary of
$A_r$ by $\chi=1-G$. Application of Minkowski functionals to the
galaxy cluster distribution can be found in \citeasnoun{kercher97}.
These
quantities have been used also as efficient shape finders by
\citeasnoun{sahnisat}.

\acknowledgments
This work was supported by the Spanish MCyT project AYA2000-2045 and
by the Estonian Science Foundation under grant 2882. Enn Saar is grateful 
for the invited professor position funded by
the Vicerrectorado de Investigaci\'on de
la Universitat de Val\`encia.

\end{document}